\documentclass[11pt]{article}
\usepackage{amsmath,amssymb,bm,epsf,epsfig,graphicx}
\input epsf.sty
\usepackage[utf8]{inputenc}
\usepackage[english]{babel}
\usepackage{amsmath}
\usepackage{latexsym}
\usepackage{amsfonts}
\usepackage{mathtools}
\usepackage{amssymb}
\usepackage{scalerel}
\usepackage{extpfeil}
\usepackage[left=3cm,right=3cm,top=3.1cm,bottom=3.1cm]{geometry}
\usepackage{cite}
\usepackage{tensor}
\usepackage{tikz}
\usepackage{tikz-cd}
\usepackage[mathscr]{euscript}
\usetikzlibrary{arrows.meta, bending}
\tikzset{C/.style={circle, minimum size=8mm,
		node contents={},
		append after command={\pgfextra{%
				\draw[-{Straight Barb[flex']}](\tikzlastnode.150) arc (450:110:2.8mm);}
	}}
}

\usepackage{hyperref}
\numberwithin{equation}{section}

\def\bra#1{\langle #1 |}
\def\ket#1{|#1 \rangle}
\def\aver#1{\left\langle\, #1 \,\right\rangle}

\def\p{\partial}

\def \be {\begin{eqnarray}}
\def \ee {\end{eqnarray}}
\def \bal {\begin{align}}
\def \eal {\end{align}}
\def \bdm {\begin{displaymath}}
\def \edm {\end{displaymath}}

\def\del {\partial}
\def\0{\nonumber}

\def \VV {{\mathbb V}}

\begin{document}
	\begingroup\allowdisplaybreaks

\vspace*{1.1cm}

\centerline{\Large \bf Closed string deformations  in open string field theory I:}\vspace{.3cm}
 \centerline{\Large \bf  Bosonic string}
\vspace*{.1cm}

\begin{center}

{\large Carlo Maccaferri$^{(a)}$\footnote{Email: maccafer at gmail.com} and Jakub Vo\v{s}mera$^{(b,c)}$\footnote{Email: jvosmera at phys.ethz.ch} }
\vskip 1 cm
$^{(a)}${\it Dipartimento di Fisica, Universit\`a di Torino, \\INFN  Sezione di Torino and Arnold-Regge Center\\
Via Pietro Giuria 1, I-10125 Torino, Italy}
\vskip .5 cm
$^{(b)}${\it Institut f\"{u}r Theoretische Physik, ETH Z\"{u}rich\\
	Wolfgang-Pauli-Straße 27, 8093 Z\"{u}rich, Switzerland}
\vskip .5 cm
$^{(c)}${\it Institute of Physics of the AS CR,\\   
Na Slovance 2,  Prague 8, Czech~Republic}

\end{center}

\vspace*{6.0ex}

\centerline{\bf Abstract}
\bigskip
This is the first of a series of three papers  on open string field theories based on Witten star product deformed with a gauge invariant open/closed coupling.  This deformation is  a tree-level tadpole which destabilizes the initial perturbative vacuum. We discuss the existence of vacuum-shift solutions which cancel the tadpole and represent a new configuration where the initial D-brane system has adapted to the change in the closed string background. As an example we consider the bulk deformation which changes the compactification radius and, to first order in the deformation, we reproduce the shift in the  mass of the open string KK modes  from the new kinetic operator after the vacuum shift. We also discuss the possibility of taming closed string degenerations with the open string propagator in the simplest amplitude corresponding to two closed strings off a disk.

\baselineskip=16pt
\newpage
\tableofcontents

\section{Introduction and summary}
 The interplay between open and closed strings is at the heart of string theory but the corresponding dynamics is in  general  not easily accessible as it requires to take into account  strong-coupling effects due to D-branes  where, as the string coupling constant grows, the perturbative  world-sheet description fades away and we loose control on the microscopic degrees of freedom. This is supposed to happen for example in the geometric transition at the core of the AdS/CFT correspondence and related scenarios where D-branes are turned into ``flux''. In principle this is not so different from what happens in QCD trying to understand low-energy strongly coupled physics  starting from the microscopic quark-gluon path integral. But to gain this conceptual and computational picture we need a space-time field theory approach. Therefore it is natural to pose the problem of open-closed physics in the framework of string field theory (SFT).

Noticeable progress  has been achieved  in the past years in the  constructions of complete RNS superstring field theories \cite{Moosavian:2019ydz, Kunitomo:2019glq, Erler:2017onq, Konopka:2016grr, Erler:2016ybs, Sen:2015uaa,  Kunitomo:2015usa}, in the understanding of the classical vacuum structure of these theories \cite{Erler:2019fye, Vosmera:2019mzw,  Erler:2019nmz, Erler:2019xof, Kudrna:2018mxa, Cho:2018nfn, Larocca:2017pbo, Kojita:2016jwe,  Kudrna:2016ack,  Maccaferri:2015cha,  Erler:2014eqa, Maccaferri:2014cpa,Kudrna:2014rya,Erler:2013wda,  Kudrna:2012re,Erler:2012qn,  Murata:2011ep, Hata:2011ke, Kiermaier:2010cf, Bonora:2010hi, Schnabl:2005gv} and in the development of perturbation theory at the quantum level  \cite{Sen:2021qdk,   Sen:2020eck, Sen:2020ruy, Sen:2020oqr, Sen:2020cef, Sen:2019qqg, Sen:2019jpm, deLacroix:2018tml, Sen:2017szq, Sen:2016qap,Sen:2016uzq, Sen:2016gqt, Pius:2016jsl, Sen-restoration, Sen:2015hha, Sen:2014dqa, Sen:2014pia}, which also provided a first-principle resolution of several worldsheet drawbacks associated with degenerations of Riemann surfaces and spurious singularities. See \cite{Erler:2019loq,Erler:2019vhl,deLacroix:2017lif} for recent reviews on SFT.

In the set of the available quantum-consistent SFTs  the only one  which is a true quantum field theory of open plus closed strings is open-closed string field theory, which exists for both  the bosonic strings \cite{Zwiebach:1990qj, Zwiebach:1997fe} and, more interestingly, for type II oriented and unoriented superstrings \cite{Moosavian:2019ydz}. An open-closed string field theory can be constructed starting from a choice of decomposition of the moduli space of punctured Riemann surfaces  in fundamental open-closed vertices and open and closed string propagators. Different decompositions give rise to different open-closed SFTs  which are however all related to each other by field redefinitions and therefore have the same physical content. In this class of theories one can in principle follow generic dynamical  changes of the closed string background and at the same time (thanks to the open string degrees of freedom) have under control the non-perturbative sectors given by D-branes on the  (dynamical) closed string background. It is therefore very important that such a complete framework exists. Unfortunately, at the same time, all orders (or even non-perturbative) computations with open-closed SFTs are  not doable because, just like closed string field theory \cite{Zwiebach:1992ie} (which is a perturbatively consistent subsector of it), the explicit world-sheet expression of the multi-string open-closed products is only known or constructible  to the first few orders (although progress is happening \cite{Firat:2021ukc, Cho:2019anu, Moosavian:2019pmd, Costello:2019fuh, Headrick:2018ncs, Moosavian:2017sev, Moosavian:2017qsp}) and it is not clear if and how non-perturbative classical solutions could be constructed.
In several physical applications one is often interested in situations where open strings are the real dynamical variables and closed strings serve as a  continuous deformation of the background on which the D-branes are initially put. In this situation the process of deformation is captured in perturbation theory by  the scattering of the deforming on-shell closed strings off the initial D-branes (including their open strings fluctuations). In this case it turns out that we actually $do$ $have$ an explicit microscopic string field theory at our disposal which is given by open string field theory based on Witten star-product, together with a simple gauge invariant open/closed coupling with an on-shell closed string state, which is often called the Ellwood invariant \cite{Hashimoto:2001sm, Gaiotto:2001ji, Ellwood:2008jh}.
 In the case of the bosonic string it is known \cite{Zwiebach:1992bw} that Witten's theory with the Ellwood invariant arises from a family of interpolating open-closed string field theories, in a limit where the local coordinate patches for the closed string collapse to  punctures and the closed strings are only allowed to be  physical (with respect to the reference background). In this limit we loose the possibility of moving in the off-shell closed string landscape but the immense advantage of this approach is that  the full bosonic worldsheet moduli space is covered with just open string propagators. 

Following \cite{Sen:2016qap, Kajiura:2001ng, Kajiura:2003ax, BS} there is interest in these days to explicitly address the computation of the effective couplings for the massless fields that one gets in string field theory by integrating out the massive excitations \cite{Erbin:2019spp,Maccaferri:2019ogq, Maccaferri:2018vwo, Asada:2017ykq, Erbin:2020eyc, Koyama:2020qfb}. In particular very recently the systematic construction of effective open-closed couplings in bosonic Witten open string field theory has been carried out in \cite{Erbin:2020eyc, Koyama:2020qfb}. 
In this and the companion   papers  \cite{OC-II, OC-III} we continue in this direction and we will also extend the open-closed effective couplings in the context of open superstring field theory. Our analysis will be at the classical level and  we will focus on the NS sector of the open superstring, but we expect that our general construction will extend  to the R sector as well as to perturbative loops.

A central physical point of our work and in particular of the present paper is to characterize the vacuum shift generated by the Ellwood invariant which acts as a tadpole for the open string\footnote{See \cite{Baba:2012cs} for a previous work attempting to characterize the vacuum shift in relation with Ellwood conjecture\cite{Ellwood:2008jh}.}.
Similarly to the shift in the 1-PI effective action discussed in \cite{Sen-restoration}, the vacuum shift solution is the response of the 
system to a tadpole, in this case a change in the closed string background, and its existence (which is in general not guaranteed) is a manifestation of the fact that the initial D-brane system is able to adapt to the closed string deformation. A BCFT analysis of this problem has been discussed in \cite{Fredenhagen:2006dn} using renormalization group analysis. In this paper we start addressing this problem in the simpler setting of the bosonic string where we show how the effective action for the massless fields allows to have under perturbative control  the vacuum shift solution and the possible obstructions to its existence. The first non-trivial effective couplings involve one open string propagator. To both ends of the propagator we can either attach an Ellwood invariant or the star product of two open strings. Therefore we get a disk with two closed string insertions (which represents a non-dynamical term in the effective action), a disk with  one closed string and  two open strings (which gives rise to a deformation of the open string kinetic term) or a disk with four open strings (which is an effective coupling for open strings).  At higher order in perturbation theory more open string propagators enter the game and more open-closed couplings can be derived.

The paper is organized as follows. In section \ref{sec:2} we address the physical problem of removing the closed string tadpole by shifting the open string vacuum to a perturbative classical solution and we review the construction of the open-closed couplings in Witten OSFT as described in \cite{Erbin:2020eyc, Koyama:2020qfb}. The possible obstructions to the existence of such a solution are precisely given by the equation of motion of the (closed-string deformed) effective action.  Starting from the OSFT on a given D-brane system we give examples of exactly marginal bulk deformations that are compatible or not with the given boundary conditions and we show how this compatibility condition is related to the vanishing of the tadpole in the effective theory where open string fields outside the kernel of $L_0$ have been integrated out.
We show that after the tadpole removal (when this is not perturbatively obstructed) both the effective action and the microscopic action will have a restored strong $A_\infty$ structure, deformed by the strength of the initial tadpole (the closed string modulus). This deformation will be reflected in the kinetic operator of the shifted theory and it will give non-trivial contributions to the mass-terms. These contributions can be computed analytically using chiral four point functions of weight 1 fields, slightly generalizing the method of \cite{BS}. In particular the closed string deformation can  give mass to initially massless open string excitations and this corresponds to D-branes moduli which are lifted by the closed string deformation. As an example we consider the fate of  a D1-brane at the self dual radius  under an  exactly marginal bulk deformation increasing the compactification radius. Initially the system has three independent massless excitations giving rise to the well-known  $SU(2)$ moduli space. We show that under the radius deformation the two $SU(2)$ currents $j^{\pm}=e^{\pm2i X}$ become tachyonic,  while the $j^3=i\sqrt2\partial X$ remains massless, in agreement with the expectation from BCFT and the string spectrum.  In section \ref{sec:closdeg} we study the first term in the cosmological constant and give a concrete proposal on how to tame the divergences due to closed string degeneration. We also point out that a divergence due to the propagation of a massless field in the closed string channel (meaning that the closed string deformation is not exactly marginal) necessarily results in an incurable logarithmic divergence which would make the theory sick, as expected. We conclude in section \ref{sec:7} with a list of further directions and open problems which can be posed already in the context of the bosonic string.

The superstring part of the story is discussed in the companion papers \cite{OC-II, OC-III}.

\section{Closed string deformations in open bosonic string field theory}\label{sec:2}
The aim of this section is to study the main aspects of our work in the simpler setting of the bosonic string. Some of the main results on the structure of the effective open/closed couplings have been already discussed in detail in \cite{Erbin:2020eyc, Koyama:2020qfb}, and here we will mostly focus on the physical aspects of the problem.
\subsection{Tadpole shift}
We start with Witten bosonic open string field theory coupled to the so-called Ellwood invariant \cite{Hashimoto:2001sm, Gaiotto:2001ji, Ellwood:2008jh}

\begin{align}
S^{(\mu)}[\Psi] = \frac{1}{2}\aver{\Psi,Q\Psi}+\frac{1}{3}\aver{\Psi,\Psi^2}+\mu\aver{\Psi, e}\,,
\label{eq:actRHS}
\end{align}
where $e\equiv V(i,-i) I $ is the identity string field $I$ with a midpoint insertion of the BRST invariant $h=(0,0)$ primary $V(z,\bar z)$, $\aver{\cdot,\cdot}$ is BPZ inner-product and Witten $*$-product is understood.
Thanks to the peculiar nature of the  midpoint insertion, this is a $\mu$-deformation of the initial theory which enjoys the same gauge invariance of the undeformed theory.  However the vacuum structure is different because the closed string provides a source term in the equation of motion
\be
Q\Psi+\Psi^2=-\mu e\label{mu-eom-bos}.
\ee
Therefore $\Psi=0$ is not a solution anymore. If we want to study the physics that is induced by the $\mu$-deformation we have  to shift the vacuum to a new equilibrium point. 
Suppose then we have found a solution $\Psi_\mu$ to \eqref{mu-eom-bos}. Now we can expand the theory around this background to find
\be
S^{(\mu)}[\Psi_{\mu}+\psi]=S^{(\mu)}[\Psi_{\mu}]+\frac12\aver{\psi, Q_\mu \psi}+\frac13\aver{\psi,\psi^2},\label{shifted-act}
\ee
where, as expected, the tadpole disappears thanks to the equation of motion
\be
\aver{\left(Q\Psi_\mu+\Psi_\mu^2+\mu e\right),\psi}=0.
\ee
The shifted action \eqref{shifted-act}  contains a non-dynamical cosmological constant
\be
S^{(\mu)}[\Psi_{\mu}]=\frac\mu2\aver{e,\Psi_\mu}-\frac16\aver{\Psi_\mu,\Psi_\mu^2},\label{cosmo}
\ee
which does not affect the local physics and a new kinetic term given by the deformed BRST operator
\be
Q_\mu\,\cdot \equiv Q\,\cdot+[\Psi_\mu,\,\cdot],
\ee
which is nilpotent as a consequence of the fact that the midpoint insertion $e$ commutes with the star product
\be
Q_\mu^2\,\cdot=\left[Q\Psi_\mu+\Psi_\mu^2,\,\cdot\right]=-\mu[e,\cdot]=0.\label{nilp-bos}
\ee
We will now assume that $\Psi_\mu$ is a vacuum shift solution which can be constructed perturbatively in $\mu$ around the $\Psi=0$ vacuum of the $\mu=0$ theory. That is, we will assume that $\Psi_{\mu=0}=0$. From the explicit form of the new action we see that, provided a vacuum shift solution $\Psi_\mu$ exists, the  difference brought in by the deformation is in the kinetic operator, which will give rise to a deformation of the physical fluctuations of the D--brane system on which the original OSFT is defined. Assuming the deforming closed string state lives in an internal CFT sector (for example a compactification), this will be perceived as a mass correction to the on-shell states which live on the D-brane. Such mass correction is explicitly given by the quadratic form
\be
\delta_\mu S[\psi]=\aver{\psi,\Psi_\mu\,\psi},
\ee
therefore the physical changes in the new background are encoded in the vacuum shift solution $\Psi_\mu$. We can search for $\Psi_\mu$  perturbatively in $\mu$\footnote{ As we will see shortly a perturbative solution is not always guaranteed to exist and it is possible, depending on the closed string deformation and the D-brane system under consideration, that a vacuum could only be found non-perturbatively. }
\be
\Psi_\mu=\sum_{\alpha=1}^\infty \mu^\alpha \psi_\alpha.
\ee
The vacuum shift equation now splits into infinite recursive equations
\be
Q\psi_1&=&- e\\
Q\psi_2&=&-\psi_1^2\\
Q\psi_3&=&-\left[\psi_1,\psi_2\right]\\
&\vdots&\0
\ee
and we can try to solve them iteratively.
The first equation for $\psi_1$ is already quite subtle. It tells us that we will be able to find a solution only if the open string state $e$ is BRST exact. Notice that this is not the statement that the closed string state $V(z,\bar z)$ that has been used to build the Ellwood state $e$ is itself BRST trivial, but rather it means that when the closed string state $V(z,\bar z)$ entering in $e$ is re-expressed in the open string channel via the bulk-boundary OPE, the generated open string states should be outside of the open string cohomology. Explicitly we can search for a solution using the Siegel gauge propagator to invert $Q$ outside of  the $L_0$ kernel. Then calling $P_0$ the projector on the kernel of $L_0$ we can write
\be
\psi_1=-\frac{b_0}{L_0}(1-P_0)\, e+ \varphi_1,
\ee
where $(1-P_0)\varphi_1=0$. We now remain with a  yet-to-be-solved equation in Ker$(L_0)$ which reads
\be
Q\varphi_1+P_0 e=0.
\ee
In general a  sufficient condition to be able to proceed  is to use a closed string deformation such that 
\be
P_0 e=0, \label{first-obs}
\ee
and we see that in this case we can simply set $\varphi_1=0$ and find a full first order vacuum shift as
\be
\psi_1&=&-\frac{b_0}{L_0}(1-P_0)\,e= -\frac{b_0}{L_0}\, e.
\ee
In fact, in the (open) zero momentum sector, the condition \eqref{first-obs} is also necessary. Indeed in such a sector the massless field $\varphi_1$ can be written in full generality as
\be
\varphi_1&=&a\del c+b_i cV_1^i,\\
Q\varphi_1&=&a c\del^2 c,
\ee
where $V^i_1$'s are generic $h=1$ matter boundary primaries. However at zero momentum, because of the symmetry of the Ellwood invariant we will have that the only possible contribution in $P_0e$ is
\be
P_0e\sim \sum_i e_i c\del c V_1^i.
\ee
So we see that $P_0 e$, if not vanishing, necessarily belongs to the open string cohomology at ghost number two and therefore the only solution to $Q\varphi_1+P_0e=0$ is $\varphi_1=Q\lambda$ and $P_0e=0$.

Since the projector condition $P_0 e=0$ will be crucial in this paper let us give two examples of simple choices of closed string deformations which obey  \eqref{first-obs} or do not. As an example of $P_0 e=0$ we can consider a D1-brane wrapping a circle of radius $R$. The radius of the circle is a closed string modulus which is controlled by the massless closed string state $V(z,\bar z)\sim c\del Y(z) \bar c\bar\del Y(\bar z)$. The corresponding Ellwood state is given by the open string field
\be
e=\frac1{2\pi i} U_1^\dagger c j(i) c  j(-i)\ket0,\label{ell-radius}
\ee
where\footnote{The imaginary normalization is needed to make $e$ a real string field. The $1/(2\pi)$ is conventionally chosen such that for a classical solution $\Psi_*$ the quantity $\aver{ e,\Psi_*}$ computes the shift in the canonically normalized disk tadpoles induced by the classical solution \cite{Ellwood:2008jh}. } we have defined the canonically normalized $U(1)$ current $j(z)\equiv i\sqrt{\frac2{\alpha'}}\del Y(z)$ and we have re-expressed $\bar j$ as $j$ according to the Neumann gluing condition $\bar j(\bar z)\to +j(z^*)$. The twist-invariant operator $U_1^\dagger$ is given by 
\be
U_1^\dagger=\exp\left(\sum_{n\geq 1}v_nL_{-2n}\right),\label{Udagger}
\ee
where the $v_n$'s are known (but unimportant in our analysis) coefficients \cite{wedges}.
The crucial quantity is the bulk-boundary  OPE
\be
j(i y)j(-i y)=\frac1{(2 i y)^2}+{\rm regular},
\ee
which does not contain any weight 1 field. Then the total matter-ghost bulk-boundary OPE is 
\be
cj(iy)cj(-iy)=-\frac{c\del c(0)}{2i y}+{\rm vanishing}.
\ee
Notice that nothing is generated at $L_0=0$.
As a consequence of this (together with the fact that the first correction  to the identity in $U^\dagger$ \eqref{Udagger} is a level two operator),  we simply find\footnote{In a generic open-closed SFT a non-twist invariant open-closed coupling would generate the level-zero BRST exact state $c\del^2 c$ which would have to be treated accordingly \cite{Sen:2020cef, Erbin:2020eyc}.}
\be
P_0e=0\,,
\ee
meaning that the radius deformation is not obstructed by the D1 brane.
A counter-example is given in the same D1-brane setting, but at the self-dual radius $R=\sqrt{\alpha'}$, where the bulk CFT has a global $SU(2)^2$ symmetry generated by the left-moving currents (decomposing $Y = Y_L+Y_R$)
\begin{subequations}
	\begin{align}
	J_L^1 &= \sqrt{2}\cos(\tfrac{2}{\sqrt{\alpha'}}Y_L)\,,\\
	J_L^2 &= \sqrt{2}\sin(\tfrac{2}{\sqrt{\alpha'}}Y_L)\,,\\
	J_L^3 &= i\sqrt{\tfrac{2}{\alpha'}} \p Y_L\,,
	\end{align}
\end{subequations}
and the analogously defined right-moving currents $J_R^1$, $J_R^2$, $J_R^3$. We then consider the BRST invariant $h=(0,0)$ closed string field $V (z,\bar z)\sim c\bar c \, J_L^1 J_R^2(z,\bar{z})
$. From a purely bulk perspective, this is an exactly marginal deformation since it can be related to the radius deformation by the global $SU(2)^2$ rotation $J_L^a \to \tilde{J}_L^a$, $J_R^a \to \tilde{J}_R^a$, where
\begin{subequations}
	\label{eq:SU2rot}
	\begin{align}
	&\tilde{J}^1_L = J^3_L\,,\qquad \tilde{J}^3_L = J^2_L\,,\qquad \tilde{J}^2_L = J^1_L\,,\\
	&\tilde{J}^1_R = J^2_R\,,\qquad \tilde{J}^2_R = J^3_R\,,\qquad \tilde{J}^3_R = J^1_R\,.
	\end{align}
\end{subequations}
The Ellwood state is given by
\be
e=\frac1{2\pi i} U_1^\dagger \,c J_L^1(i)\, c  J_R^2(-i)\ket0.
\ee
In this case the bulk-boundary OPE generates the marginal open string field $j$
\be
J_L^1(iy)\, 
J_R^2(-iy)=
i\frac{j(0)}{2iy} +{\rm regular}\label{violation}
\ee
and the projector condition is violated
\be
P_0e=c\del cj(0)\ket0\neq0.
\ee
Notice that $P_0e$ is in the ghost number 2 open string cohomology and therefore the equation $Q\varphi_1+P_0e=0$ does not admit a solution. This is therefore a true obstruction for the vacuum shift which is telling us that the D1-brane at the self-dual radius cannot adapt to this bulk deformation. Physically, we have the following interpretation for this obstruction. The boundary state $\| \text{D1}\rangle\!\rangle$ for the D1 brane (with trivial Wilson line) satisfies the gluing conditions
\begin{subequations}
	\begin{align}
	\big[(J_L^1)_n+(J_R^1)_{-n}\big]\|\text{D1}\rangle\!\rangle &=0\,,\\
	\big[(J_L^2)_n+(J_R^2)_{-n}\big]\|\text{D1}\rangle\!\rangle &=0\,,\\
	\big[(J_L^3)_n+(J_R^3)_{-n}\big]\|\text{D1}\rangle\!\rangle &=0\,.
	\end{align}
\end{subequations}
Performing the global $SU(2)^2$ rotation \eqref{eq:SU2rot} which maps the $J_L^1 J_R^2$ bulk deformation to the radius deformation, these gluing conditions become
\begin{subequations}
	\begin{align}
	\big[(J_L^3)_n+(J_R^2)_{-n}\big]\widetilde{\|\text{D1}\rangle\!\rangle} &=0\,,\\
	\big[(J_L^1)_n+(J_R^3)_{-n}\big]\widetilde{\|\text{D1}\rangle\!\rangle} &=0\,,\\
	\big[(J_L^2)_n+(J_R^1)_{-n}\big]\widetilde{\|\text{D1}\rangle\!\rangle} &=0\,,
	\end{align}
\end{subequations}
where the boundary state $\widetilde{\|\text{D1}\rangle\!\rangle}$ now clearly does not encode neither Neumann, nor Dirichlet boundary conditions: instead, it describes a conformal brane at an intermediate point in the $SU(2)$ moduli space \cite{Gaberdiel:2001xm,Gaberdiel:2001zq}, for which the radius deformation is known to induce a boundary RG flow \cite{Fredenhagen:2006dn}. Hence, on the grounds of the global $SU(2)^2$ invariance of the free-boson CFT at the self-dual radius, it follows that the $J_L^1 J_R^2$ bulk deformation should be obstructed by the D1 brane, in agreement with our SFT result.

Going to higher orders in $\mu$ and setting to zero the possible $L_0=0$ contribution $\varphi_\alpha$ to the solution $\Psi_\mu$, we find an infinite set of constraints for the tadpole $e$
\be
P_0e&=&0\0\\
P_0(h e)^2&=&0\label{obstruct}\\
P_0[he,h(he)^2]&=&0,\0\\
&\cdots&\0
\ee
where we have defined the propagator
\be
h\equiv\frac{b_0}{L_0}(1-P_0).\label{homotopy}
\ee
These conditions which are (open) string-field-theoretic in nature can be considered as sufficient conditions for the existence of  a deformation of a given worldsheet boundary condition generated by a given bulk deformation. Looking at their structure, we realize that they are setting to zero the amplitudes involving arbitrary number of deforming closed strings and a single massless (i.e. in the kernel of $L_0$) open string. As we are now going to see, these conditions are just stating the absence of a tadpole in the effective theory for the massless fields $\varphi$. 
\subsection{Effective action and open-closed couplings}

 The space-time meaning of the conditions \eqref{obstruct} can be better understood in terms of the effective action for the massless fields, as discussed in \cite{Erbin:2020eyc}.  We can classically integrate out all the fields outside of the kernel of $L_0$ and remain with an effective theory describing effective interactions between massless open strings and the closed string entering the Ellwood invariant. To do so we split the total string field using the projector on the kernel of $L_0$
 \be
 \Psi=P_0\Psi+(1-P_0)\Psi=\varphi+R,
 \ee
and we integrate out classically the massive string field $R$ as a function of the massless one $\varphi$. The $R$-equation is simply the $(1-P_0)$-projected EOM and reads
\be
(1-P_0)[Q\Psi+{\cal J}_\mu(\Psi)]=QR+(1-P_0){\cal J}_\mu(\varphi+R)=0,
\ee
where we have defined the interacting part of the EOM as
\be
{\cal J}_\mu(\Psi)=\Psi^2+\mu e={\cal J}_0(\Psi) +\mu e.\label{J-mu}
\ee
We can fix the gauge $hR=0$, where  the propagator $h$ has been defined in \eqref{homotopy}
 and find an equivalent ``integral equation'', by acting on the massive EOM with $h$\footnote{The out-of-gauge massive equations that we will miss in this way are automatically accounted for by the massless equations of the effective action as discussed extensively in \cite{Erbin:2020eyc}.}
 \be
 R=-h{\cal J}_\mu(\varphi+R),
 \ee
 that is
 \be
 \Psi(\varphi)=\varphi-h{\cal J}_\mu(\Psi(\varphi)).\label{def-poppo}
 \ee
 We can easily solve this equation assuming we have already solved the corresponding equation {\it without} the closed string deformation. Let $\Psi_0(\varphi)$ be such a solution
 \be
 \Psi_0(\varphi)=\varphi-h{\cal J}_0(\Psi_0(\varphi)),\label{poppo}
 \ee
 which can be perturbatively expressed as
\be
 \Psi_0(\varphi)=\varphi-h(\varphi^2)+h[\varphi,h (\varphi^2)]+ O(\varphi^4).\label{sol-poppo}
\ee
It is important to notice that although equation \eqref{poppo} is originally understood for $\varphi\in{\rm Ker}(L_0)$, equation \eqref{sol-poppo} makes sense for generic $\varphi=\chi$, not necessarily in the kernel of $L_0$. In other words, given
\be
\Psi_0(\chi)=\chi-h(\chi^2)+h[\chi,h (\chi^2)]+ O(\chi^4),
\ee
this formally
 provides a solution to 
\be
 \Psi_0(\chi)=\chi-h{\cal J}_0(\Psi_0(\chi)),\label{poppo-chi}
\ee
for generic $\chi$. 
This observation is useful to solve the deformed equation \eqref{def-poppo}. Indeed using \eqref{J-mu} we readily find that  \eqref{def-poppo} can be re-written as
\be
\Psi(\varphi)=(\varphi-\mu h e)-h{\cal J}_0(\Psi(\varphi)).
\ee
Then it is immediate to verify that the  solution to \eqref{def-poppo} is given by
 \begin{align}
 \Psi(\varphi)&=\Psi_0(\varphi-\mu h e)\label{tet-field}\\
 &=(\varphi-\mu h e)-h((\varphi-\mu h e)^2)+h[(\varphi-\mu h e),h ((\varphi-\mu h e)^2)]+ O((\varphi-\mu h e)^4).\0
 \end{align}
The $\mu$-deformed effective action is finally obtained by substituting \eqref{tet-field} into the original action
\be
S_{\rm eff}^{(\mu)}[\varphi]&=&S^{(\mu)}[\Psi(\varphi)]=S^{(0)}[\Psi_0(\varphi-\mu he)]+\mu\aver{e,\Psi_0(\varphi-\mu he)}.
\ee
As shown in \cite{Erbin:2020eyc, Koyama:2020qfb} this can be explicitly written in the following form
\be
S_{\rm eff}^{(\mu)}[\varphi]=S^{(\mu)}[\Psi(0)]+ \sum_{k=0}^\infty\sum_{\alpha=0}^{\infty}\frac{\mu^\alpha}{k+1}\omega\left(\varphi,  n_{k\alpha}\left(\varphi^{\otimes k}\right)\right),\label{Seffmu}
\ee
where we have used a suspended notation which makes explicit the (weak)  $A_\infty$ structure. In particular we have defined the symplectic form 
\be
\omega(\varphi_1,\varphi_2)=\langle \omega|\varphi_1\otimes\varphi_2=-(-1)^{d(\varphi_1)}\aver{\varphi_1,\varphi_2}.
\ee
where the degree $d(\varphi)$ is given by the ghost number augmented by one (mod 2).
The open-closed couplings $n_{k\alpha}$ are given as
\begin{subequations}\label{eq:scoop1}
	\begin{align}
n_{01}&=P_0 e\,,\\
n_{k\alpha}(\varphi^{\otimes k}) &=\hspace{-2mm} \sum_{\substack{{l_1,\ldots,l_\alpha\geq 0}\\ \sum_{i=1}^{\alpha+1} l_i = k}} (-1)^\alpha\tilde{m}_{k+\alpha}(\varphi^{\otimes l_1},h_0e,\varphi^{\otimes l_2},h_0e,\ldots,\varphi^{\otimes l_\alpha},h_0e,\varphi^{\otimes l_{\alpha+1}})\,\,,
\end{align}
\end{subequations}
where the last line is valid for $(k,\alpha)\neq(0,1)$. As discussed above they are constructed using the effective purely open string products $\tilde m_k$  which are explicitly given by (see, for example, section 3.1 of \cite{Erbin:2020eyc})
\be
\tilde m_2(\varphi_1,\varphi_2)&=&P_0 m_2(\varphi_1,\varphi_2)\\
\tilde m_3(\varphi_1,\varphi_2,\varphi_3)&=&-P_0\left[m_2\left(h m_2(\varphi_1,\varphi_2),\varphi_3\right) + m_2\left(\varphi_1,h m_2(\varphi_2,\varphi_3)\right)\right]\0\\
&\vdots&,\0
\ee
where the 2-product $m_2$ is the suspended version of Witten star product
\be
m_2(\varphi_1,\varphi_2)=(-1)^{d(\varphi_1)}\varphi_1\,\varphi_2.
\ee
Looking at \eqref{Seffmu}, besides the  non-dynamical cosmological constant  $S^{(\mu)}(\Psi(0))$, for $k=0$ we find the effective tadpole
\be
\sum_{\alpha=1}^{\infty}(-\mu)^\alpha\omega\left(\varphi,n_{0\alpha}\right)=\sum_{\alpha=1}^{\infty}(-\mu)^\alpha\omega\left(\varphi,\tilde m_\alpha\left((he)^{\otimes\alpha}\right)\right).
\ee
Notice that this tadpole contains, order by order in $\mu$, all the conditions \eqref{obstruct} which guarantee the existence of a vacuum shift in the full theory. In this paper we will always consider situations where this massless tadpole vanishes, so that $\varphi=0$ is a solution to the effective equations of motion.  
Starting from $k=2$ we see that the terms in the effective action correspond to tree-level amplitudes of $P_0$-projected open strings and physical closed strings. 
\subsection{Example: the radius deformation}\label{sec:radius}
Consider a D1-brane wrapping a circle (with coordinate $Y$) at the self-dual radius $R=\sqrt{\alpha'}$. The compactification radius is a closed string modulus which is controlled by the exactly marginal bulk operator $\del Y(z)\bar\del Y(\bar z)$. We thus consider the OSFT on the D1-brane deformed by the Ellwood invariant  discussed in \eqref{ell-radius}
\be
e=\frac1{2\pi i} U_1^\dagger c j(i) c  j(-i)\ket0\0.
\ee
Since we have $P_0e=0$ the tadpole can be removed to the first order in $\mu$ (in fact we expect that all obstructions \eqref{obstruct} vanish in this case just because a D1-brane trivially exists for all compactification radii), therefore it is interesting to see how the physical open string spectrum is deformed. Before the deformation,  at the self-dual radius, there are three massless states
\be
\varphi_0&=&\phi_0 cj\\
\varphi_{\pm}&=&\phi_{\pm}ce^{\pm \frac{2i}{\sqrt{\alpha'}} Y},
\ee
where the $\phi's$ are the spacetime fields (for convenience taken at zero momentum).
These three massless fields are  the Goldstone bosons of the the well-known $SU(2)$ D-branes moduli space \cite{Gaberdiel:2001xm,Gaberdiel:2001zq}.
When we switch on the deformation, looking at \eqref{Seffmu}, we find the following mass corrections\footnote{An analogous computation for the shift in the mass spectrum of the closed-string KK modes has been presented in \cite{Sen:2019jpm}.}
\be
\frac12\phi_0  m^2_0\phi_0&=&\mu\bra{e}\frac{b_0}{L_0}\ket{\varphi_0*\varphi_0}=\mu\omega\left(m_2(\varphi_0,\varphi_0),he\right)\\
\phi_+m^2_{\pm}\phi_-&=&\mu\bra{e}\frac{b_0}{L_0}\ket{\varphi_+*\varphi_-}+(+\leftrightarrow-)=\mu\omega\left(m_2(\varphi_+,\varphi_-),he\right)+(+\leftrightarrow-).
\ee
These are OSFT amplitudes which can be easily evaluated  using the BRST structure to flatten-down the world-sheet diagrams to the UHP without explicit need of a Schwarz-Christoffel map \cite{giddings}. Following \cite{BS} and \cite{wedges} we are interested in computing
\be
(2\pi i)\bra{e}\frac{b_0}{L_0}\ket{\varphi_a*\varphi_b}=\bra0 cj(i)cj(-i)U_1\frac{b_0}{L_0}U_3^\dagger\varphi_a(\sqrt3)\varphi_b(-\sqrt3)\ket0=(*),
\ee
where $\varphi_a(z)=cj_a(z)$, being $j_a$ one of the three $SU(2)$ boundary currents.
On the left of $U_1$ we insert the Hodge-Kodaira resolution of the  identity 
\be
Qh+hQ+P_0=1.
\ee
Then we notice that  $Q h$ does not contribute thanks to BRST invariance and  $P_0$ can also be dropped because of $P_0 (cj(i)cj(-i))=0$. This leaves us with
\be
(*)&=&\bra0 cj(i)cj(-i) h QU_1hU_3^\dagger\varphi_a(\sqrt3)\varphi_b(-\sqrt3)\ket0\nonumber\\
&=&\bra0 cj(i)cj(-i) h U_1QhU_3^\dagger\varphi_a(\sqrt3)\varphi_b(-\sqrt3)\ket0\0\\
&=&\bra0 cj(i)cj(-i) h U_1 U_3^\dagger\varphi_a(\sqrt3)\varphi_b(-\sqrt3)\ket0\nonumber\\
&=&\bra0 cj(i)cj(-i) h U_4^\dagger U_{\frac43}\varphi_a(\sqrt3)\varphi_b(-\sqrt3)\ket0=(**)\0
\ee
where we have made use of the gluing theorem for wedge states \cite{wedges}
\be
U_rU_s^\dagger=U^\dagger_{2+\frac2r(s-2)}U_{2+\frac2s(r-2)},\label{gluing-wedges}
\ee
and we have also used (in going from second to third line) that 
\be
P_0U_1^\dagger h  \,cj(i)cj(-i) \ket0=0,
\ee
by direct inspection of the $cj$-$cj$ OPE (which is also responsible for $P_0e=0$).
Finally applying the conformal transformation
\be
U_r \phi(z)\ket0&=&f_r\circ \phi(z)\ket0,\\
f_r(z)&=&\tan\left(\frac2r \tan^{-1}z\right),\label{wedgemap}
\ee
we get a pure UHP correlator
\be
(**)&=&\bra0 cj(i)cj(-i) \frac{b_0}{L_0} \varphi_a(1)\varphi_b(-1)\ket0\0\\[1mm]
&=&\bra0 cj(i)cj(-i) \frac{b_0}{L_0+\epsilon} \varphi_a(1)\varphi_b(-1)\ket0{\Bigg |}_{\epsilon=0}\nonumber\\
&=&\int_0^1\frac{dt}{t}t^\epsilon\left\langle cj(i)cj(-i) b_0\left( cj_a(t)cj_b(-t)\right)\right\rangle{\Bigg |}_{\epsilon=0}\0\\
&=&\int_0^1dt\,t^\epsilon\left\langle cj(i)cj(-i)\left(c(t)+c(-t)) j_a(t)j_b(-t)\right)\right\rangle{\Bigg |}_{\epsilon=0}\0\\
&=&\int_0^1dt\,t^\epsilon\,4i(1+t^2)\left\langle j(i)j(-i) j_a(t)j_b(-t)\right\rangle{\Bigg |}_{\epsilon=0}\,,\nonumber
\ee
whose actual value depends on the matter four-point function $\aver{jjj_aj_b}$. Notice that to correctly treat the ``tachyon'' divergence  due to possible propagation of fields with total weight $h<0$ we have set
\be
\frac{b_0}{L_0}\to\frac{b_0}{L_0+\epsilon}{\Bigg |}_{\epsilon=0}=b_0\int_0^1 \frac {dt}t\, t^\epsilon\, t^{L_0}{\Bigg |}_{\epsilon=0}\label{tame}
\ee
with the understanding to perform the computation in the region of $\epsilon$ where the integral is convergent. At the end we analytically continue to $\epsilon\to0$. As discussed in section 3 of \cite{Larocca:2017pbo} (in a related conformal frame), this has the effect of setting $\frac1{L_0}$ equal to $-\frac1{|h|}$ when it acts on a $h<0$ state.\footnote{This is equivalent to the prescription discussed in \cite{Sen:2019jpm}, we thank Ashoke Sen for a discussion.}
This gives the result
\be
\omega\left(m_2(\varphi_a,\varphi_b),he\right)=\frac{2\phi_a\phi_b}{\pi}\int_0^1dt\,t^\epsilon(1+t^2)\left\langle j(i)j(-i) j_a(t)j_b(-t)\right\rangle{\Bigg |}_{\epsilon=0}.
\ee
The actual value of this mass term depends on the matter chiral four point function $\langle j jj_aj_b\rangle$. 
Let us start with the potential mass correction for $\phi_0$ (which we expect to vanish since we know that this is an exactly marginal boundary field at all radii). The chiral four point function gives
\be
\left\langle j(i)j(-i) j(t)j(-t)\right\rangle=	\frac1{(t-i)^4}+\frac1{(t+i)^4}-\frac1{16 t^2}.
\ee
Evaluating the integral in the convergence region and setting $\epsilon=0$ at the end we find
\be
\omega\left(e,hm_2(\varphi_0,\varphi_0)\right)=\frac{2\phi_0\phi_0}{\pi}\left(-\frac i2+\frac i2-\frac{\epsilon}{8(\epsilon^2-1)}{\Bigg |}_{\epsilon=0}\right)=0.
\ee
This is the expected result:  the field $cj$  remains massless as it corresponds to the Wilson line deformation which  exists for any radius. Notice that the correct regularization of the tachyon divergence is crucial for getting the correct physics. For example one could think that a replacement $t^\epsilon\to (t f(t))^\epsilon$  would give the same result if $f(t)$ is regular at the origin, but this is not true (as the reader can easily check by doing for example $t^\epsilon\to\left(\frac t{1+a t}\right)^\epsilon$). This regularization is guaranteed to work only  if the integration variable is correctly related to the Schwinger parameter of $\frac1{L_0}$. Analogous issues are discussed in \cite{Sen:2020eck, Sen:2019jpm} for the analogous regularization used there.

Let us now analyze the fate of the mass term for the other two marginal directions $\varphi_{\pm}$. In this case we have
\be
\left\langle j(i)j(-i) e^{ \frac{2i}{\sqrt{\alpha'}}  Y}(t)e^{- \frac{2i}{\sqrt{\alpha'}}Y}(-t)\right\rangle=	\frac1{2t^2}\left(-\frac18+\frac{4t^2}{(1+t^2)^2}\right).
\ee
Evaluating the mass correction this time we find
\be
\omega\left(e,hm_2(\varphi_+,\varphi_-)\right)=\frac{2\phi_+\phi_-}{\pi}\int_0^1dt\,t^\epsilon\left(-\frac1{16 t^2}-\frac{1}{16}+\frac2{1+t^2}\right){\Bigg |}_{\epsilon=0}=\frac12 \phi_-\phi_+.
\ee
This is correctly saying that the moduli $\phi_\pm$ are lifted by the radius  deformation and to this order in $\mu$ we find
\be
\alpha' m^2_\pm=\mu.\label{SFT-mass}
\ee
To relate $\mu$ with the physical change in the radius $R$ we can look at the mass formula of the KK tower  
\be
\alpha' m^2=\alpha'\left(\frac nR\right)^2-1,
\ee
which is precisely massless at the self-dual radius $R=\sqrt{\alpha'}$ when $n=\pm1$.
Under a change in the radius $R\to R+\delta R$ we find
\be
\alpha'\delta m^2=-2\alpha'\left(\frac nR\right)^2\frac{\delta R}{R},\label{shift-KK}
\ee
which at the self dual radius for $n=\pm1$  gives 
\be
\alpha'\delta m_\pm^2=\alpha' m_\pm^2=-2\frac{\delta R}{\sqrt{\alpha'}}\,.
\ee
so that
\begin{align}
\mu|_{R=\sqrt{\alpha'}} = -2\frac{\delta R}{\sqrt{\alpha'}}\,.
\end{align}
This is correctly saying that these moduli become tachyonic as the radius increases, providing  the RG direction for the D1-brane to flow to the less energetic D0-brane. To derive the relation between $\mu$ and $\delta R$ for general $R$, let us generalize our computation for the mass term correction to any KK state at generic radius.
This time, in order to stay inside the cohomology and to be able to compute the amplitude without the Schwarz-Christoffel map we have to introduce space-time momentum $k_\mu$ along $p$ non-compact Neumann directions in addition to the KK modes as
\be
\varphi_n^{\pm}(k,z)=\phi_n^{\pm}(k)\, c\,e^{2i k\cdot X}\, e^{\pm2i \frac nR Y}(z){\Bigg |}_{k^2=1/\alpha'-\frac{n^2}{R^2}} ,
\ee
which is a $h=0$ BRST invariant boundary primary.
Because the conformal properties are exactly the same as in the zero momentum case at the self dual radius, the mass computation proceeds unaffected and we are left with
\begin{align}
&\omega\left(m_2(\varphi^+_n(k),\varphi^-_n(k')),he\right)=\0\\
&=\frac{2\phi^+_n(k)\phi^-_n(k')}{\pi}\int_0^1dt\,t^\epsilon(1+t^2)\left\langle j(i)j(-i) e^{2i k\cdot X}\, e^{2i \frac nR Y}(t)e^{2i k'\cdot X}\, e^{-2i \frac nR Y}(-t)\right\rangle{\Bigg |}_{\epsilon=0}.
\end{align}
Using $ k^2= k'^2=1/\alpha'-\frac{n^2}{R^2}$ the matter four-point function is easily evaluated as
\be
\left\langle j(i)j(-i) e^{2i k\cdot X}\, e^{2i \frac nR Y}(t)e^{2i k'\cdot X}\, e^{-2i \frac nR Y}(-t)\right\rangle=\frac1{2t^2}\left(-\frac18+\frac{2\left(\sqrt{2\alpha' }\frac nRt\right)^2}{(1+t^2)^2}\right)\delta(k+k')\0
\ee
and evaluating the integral we find  the mass correction 
\begin{align}
&\omega\left(m_2(\varphi^+_n(k),\varphi^-_n(k')),he\right)=\0\\
&\hspace{2cm}=\frac{2\phi^+_n(k)\phi^-_n(k')}{\pi}\delta(k+k')\int_0^1dt\,t^\epsilon\left(-\frac1{16 t^2}-\frac{1}{16}+\frac{\left(\sqrt{2\alpha' }\frac nR\right)^2}{1+t^2}\right){\Bigg |}_{\epsilon=0}\0\\
&\hspace{2cm}=\frac{\left(\sqrt{2\alpha' }\frac nR\right)^2}4\phi^+_n(k)\phi^-_n(k')\delta(k+k')\,.
\end{align}
Therefore from the SFT computation we find
\be
\alpha' \delta m^2_n=\frac{\left(\sqrt{2\alpha' }\frac nR\right)^2}2\,\mu= \alpha'\left( \frac nR\right)^2\,\mu\,.
\ee
By comparing with the shift derived from the KK spectrum \eqref{shift-KK} we can finally  relate the SFT deformation $\mu$ to the change in the radius $\delta R$ as
\be
\mu=-2\frac{\delta R}{R}\,.
\ee
Notice that  $\mu$ this depends logarithmically on the variation of the compactification radius.
This relation will receive corrections at higher powers of $\mu$ or $\delta R$ which will be captured by similar amplitudes with two physical open strings and an arbitrary number of closed strings as given by the open/closed effective vertices we have derived.
\section{On closed string degeneration}\label{sec:closdeg}

In this section we would like to make some preliminary yet instructive steps for  treating  closed string degeneration inside Witten OSFT. When the closed strings have generic momenta this is not necessary but at zero momentum (which is the case for the  background deformations we are interested in) the open-closed couplings we have derived will produce, upon direct computation, naively divergent quantities (analogous to the open strings divergences discussed in \cite{BS, Larocca:2017pbo, Sen:2020cef}) to which we have to be able to assign a definite finite value, if the theory is consistent. In this case working at finite momentum is not really helping because the zero momentum limit of string amplitudes is in general not well-defined. So we search for a background-independent regulator, as in the case of open string degeneration \eqref{tame}. To solve this problem at all orders in perturbation theory is  beyond the scope of this paper and here we will only consider the first non-trivial diagram where this problem arises, namely the first contribution to the cosmological constant $\aver{e,he}$. 

\subsection{Two massless closed strings off a disk at zero momentum}
\label{app:A}

We start  to compute the quantity $\aver{e,h e}$ for the radius deformation \eqref{ell-radius} \footnote{This amplitude in Witten OSFT has been also studied in \cite{Takahashi:2003kq,Garousi:2004pi}.}
\be
\Gamma=\aver{e,he}=-\frac1{4\pi^2}\bra0 V(i,-i)U_1\frac{b_0}{L_0+\epsilon}U_1^\dagger V(i,-i)\ket0.
\ee
Using the same flattenization method reviewed in section \ref{sec:radius}, the fact that $P_0e=0$ and that the wedge maps $f_r$ \eqref{wedgemap} don't move the midpoint $z=\pm i$, the Witten diagram can be simply re-written as\footnote{We infinitesimally deform $U_1\to U_{1+\tilde\epsilon}$ and use the fact that $\bra0 V(i,-i)U_r\frac{b_0}{L_0}U_s^\dagger V(i,-i)\ket0$ doesn't depend on $r$ and $s$ when $V$ is a $h=(0,0)$ primary. Using \eqref{gluing-wedges} we obtain wedge functions $f_{r,s}$ with $r,s<1$ which have singularities inside the unit disk, but this is not a problem for the correlator at hand as the only insertions are at the midpoint.}
\be
\Gamma&=&-\frac1{4\pi^2}\bra0 V(i,-i)\frac{b_0}{L_0+\epsilon} V(i,-i)\ket0\0\\
&=&\frac{1}{\pi^2}\int_0^1dt\, t^{\epsilon} (1-t^2) \aver{\VV^{\rm matter}(i,-i)\VV^{\rm matter}(it,-it)}.\label{ehe}
\ee
Focusing on the radius deformation $\VV(z,z^*)=j(z)j(z^*)$ and computing the correlator between the abelian currents $j$ we end up with
\be
\Gamma=\frac{1}{\pi^2}\int_0^1dt\, t^{\epsilon} (1-t^2) \left(\frac1{16 t^2}+\frac{1}{(1-t)^4}+\frac{1}{(1+t)^4}\right).\label{gamma}
\ee
We see that the first two terms in parenthesis have a divergence respectively  in the open ($t\to0$) and in the closed ($t\to1$)  degeneration channels. However the standard definition of the open string propagator can only cure the open channel, with the $t^\epsilon$. 
Our aim is then to find a consistent regularization for the second term. In this regard  we notice that in \cite{Sen:2020oqr} the same C-C disk-amplitude (but in a different physical background and, importantly, at generic momentum) has been computed. There, to confront with the closed string degeneration, the replacement rule has been used
\be
\int^1 dt(1-t)^{z-1} \to \int^{1-a}(1-t)^{z-1}+\frac{a^z}{z+i\varepsilon},\label{sen-reg}
\ee
where $0<a<1$ and $\varepsilon$ has to be sent to zero.
Here we can use the effectively equivalent  prescription
\be
\int^1 dt (1-t)^{z-1}\to\int^1 dt (1-t)^{z-1+\epsilon_{\rm closed}},
\ee
where the integral is done in the $\epsilon_{\rm closed}$ convergence region and $\epsilon_{\rm closed}\in \mathbb{C}$ is sent to zero at the end. 
Provisionally we will use this prescription and see where it leads. A rather immediate issue we notice is that we have to use either $t^{\epsilon_{\rm open}}$ or $(1-t)^{\epsilon_{\rm closed}}$. If we use them together as $\left(t^{\epsilon_{\rm open}}(1-t)^{\epsilon_{\rm closed}}\right)$, the two regularizations interfere with each other by creating ``spurious'' poles which make the limit $(\epsilon_{\rm open},\epsilon_{\rm closed})\to(0,0)$ undefined. In fact this ambiguity is just a glimpse of what happens by computing the amplitude \eqref{ehe}  giving finite massless momentum to the currents $j$, going to the convergence region and then attempting at continuing to zero momentum. The limit does not appear to exist. We will instead use $t^{\epsilon_{\rm open}}$ for the term affected with open string degeneration and $(1-t)^{\epsilon_{\rm closed}}$ for the one affected with closed string degeneration. With this prescription (to be justified --and in fact improved-- below), we can straightforwardly
evaluate the three contributions \eqref{gamma}
\be
\int_0^1 dt \, t^{\epsilon_{\rm open}}\,  \frac{1-t^2}{16 t^2}=-\frac1{8(1-\epsilon_{\rm open}^2)}\,&\to&\,-\frac1{8},\\\
\int_0^1 dt \, (1-t)^{\epsilon_{\rm closed}}\,  \frac{1+t}{(1-t)^3}=-\frac{\epsilon_{\rm closed}}{2-2\epsilon_{\rm closed}+\epsilon_{\rm closed}^2}\,&\to&\,0,\label{wrong-reg}\\
\int_0^1 dt \,   \frac{1-t}{(1+t)^3}&=&\frac14.
\ee
The regularization of the open string degeneration is the usual one and it does not require more explanation. Let us instead motivate the  regularization of the closed string divergence in the second term.  
The regulating $(1-t)^{\epsilon_{\rm closed}}$ is supposed to have the same origin of $t^{\epsilon_{\rm open}}$, but when we look at the process in the closed string channel \cite{Sen:2020oqr}. In this channel we should see two closed strings interacting and connecting  to the boundary state via the closed string propagator 
\be
\frac{b_0^+}{L_0^++\epsilon_{\rm closed}}=\int_0^1 dx\,   x^{\epsilon_{\rm closed}-1}\,b_0^+\,x^{L_0^+}.\label{closed-prop}
\ee
The $x$ parameter is then expected to be ``equivalent'' to $(1-t)$
\be
x=\alpha(1-t)+O(1-t)^2,\quad \alpha>0
\ee
giving a  justification for the regularization \eqref{wrong-reg}. 
However, since we are at zero momentum,  this regularization is very delicate and in general if we exchange $(1-t)^\epsilon$ with $\left((1-t)f(t)\right)^\epsilon$  the result changes, as the reader can easily check in \eqref{wrong-reg}. In fact, just as for the open string degeneration analyzed in the previous section,  in order to use this regularization without concern we have to correctly relate the integration variable to the Schwinger parameter of the closed string propagator $\frac{b_0^+}{L_0+\epsilon_{\rm closed}}$. Concretely, we have to relate the $t$ variable in \eqref{wrong-reg} with $x$ in \eqref{closed-prop}. To do this we perform the $SL(2,C)$ (modulus-dependent) conformal transformation
\be
f_t(z)=\frac{\sqrt t+i z}{\sqrt t- iz},\label{map}
\ee
in such a way that the two closed strings in \eqref{ehe} will be mapped to the symmetric positions
\be
it&\to& f_t(it)=\frac{1-\sqrt{t}}{1+\sqrt{t}}\equiv x\label{change}\\
i&\to& f_t(i)=\frac{-1+\sqrt{t}}{1+\sqrt{t}}= -x.
\ee
Now, using  \eqref{change} as a change of integration variable, we can prove the following important relation
\be
 \langle 0 |\,V_1(i,-i)\frac{b_0}{L_0}V_2(i,-i)\,|0\rangle&=&-\langle B|\, \frac{b_0^+}{L_0^+} \,V_1(-1,-\bar1)V_2(1,\bar1)\,|0\rangle,\label{open-closed}\\
 &\updownarrow&\0\\
 \int_0^1 \frac{dt}{t} \langle 0 |\,V_1(i,-i)\,b_0V_2(it,-it)\,|0\rangle&=&\int_1^0\frac{dx}x\langle B|\, b_0^+ \,V_1(-x,-\bar x)V_2(x,\bar x)\,|0\rangle
\ee
where $\langle B|$ is the boundary state of the BCFT at hand and the brackets in the $rhs$ are computed in the closed string Hilbert space.  This is proven in appendix \ref{app:2}. Notice that $t=0$ corresponds to $x=1$ and vice versa.  This relation is algebraically true but notice that the left hand side involving 
open string propagation is not naturally protected against closed string degeneration $t\to1$ (when the strip of the open string propagator becomes very short), while the right hand side, involving closed string propagation, is not naturally protected against open string degeneration $x\to1$ (when the tube of the closed string propagator becomes very short). Therefore, when we are near to closed string degeneration, we have to think of $\Gamma$ as the  right hand side of \eqref{open-closed} and we have to regulate the collision with $x^{\epsilon_{\rm closed}}$ instead of the naively equivalent  $(1-t)^{\epsilon_{\rm closed}}$.\footnote{In \cite{Sen:2020oqr} this subtlety is not important because the amplitude is computed  in a region of analyticity of the external momenta. But zero momentum is not a point of analyticity and therefore we have to know in which direction to approach the singular point.}
Therefore the correct regularization in \eqref{wrong-reg} should be obtained by replacing
\be
(1-t)^{\epsilon_{\rm closed}}\to x(t)^{\epsilon_{\rm closed}}=\left(\frac{1-\sqrt{t}}{1+\sqrt{t}}\right)^{\epsilon_{\rm closed}}=\left(\frac14 (1-t)+O(1-t)^2\right)^{\epsilon_{\rm closed}}.\label{x(t)}
\ee
In the $\epsilon_{\rm closed}\to 0$ limit the $\frac14$ coefficient will be unimportant but the subleading contributions are not.
Indeed now in \eqref{wrong-reg}  we find
\be
\int_0^1 dt \, x(t)^{\epsilon_{\rm closed}}\,  \frac{1+t}{(1-t)^3}=-\frac1{8-2{\epsilon^2_{\rm closed}}}\,&\to&\,-\frac18,
\ee
and summing up the other contribution this time we get
\be
\Gamma=\aver{e,he}=0,\quad\quad\textrm {(radius deformation)}.
\ee
Notice that our argument is  naturally based on the simple relation \eqref{open-closed} which is a fact concerning an on-shell amplitude and as such we expect it to be independent of the off-shell details of SFT.
That said, it would be obviously desirable  to have an independent check of this derivation, for example a computation in open-closed SFT (see also comments in the conclusions \label{sec:7}).
\subsection{Massless  divergence and violation of bulk marginality}
\label{app:B}
From a physical point of view we expect that deforming OSFT with a closed string tadpole should only be consistent when the closed string is not just on-shell, but also exactly marginal. A necessary condition for a matter bulk field  $\VV$ to be exactly marginal is that  in the $\VV$-$\VV$ OPE no weight $(1,1)$ field is produced. Therefore we expect that something pathological should happen if a bulk  massless field is produced in the collision of the two Ellwood invariants. In fact, we can see this pathology very clearly  in $\aver{e,he}$.
Although the regularization of the closed string tachyon divergence we have discussed in the previous subsection is very sensitive to the details of the function $x(t)$ \eqref{x(t)}, the  breakdown of the computation due to massless propagation in the closed string channel is on the other hand very clear and unambiguous. A massless divergence (i.e the propagation of a $(1,1)$ matter field) will correspond to a $\frac{A}{1-t}$ term in the integrand of \eqref{ehe}. In this case it is not difficult to see that, independently of the regulating function \eqref{x(t)}, computing the integral in the $\epsilon$ convergence region will always produce a $\frac 1{\epsilon_{\rm closed}}$ result, with a coefficient that is independent of the regulating function
\be
\int_0^1 dt {\Big(}a(1-t)+O(1-t)^2{\Big)}^{\epsilon_{\rm closed}} \frac A {1-t}=\frac{A}{\epsilon_{\rm closed}}+O(1).
\ee
Differently from the tachyon divergence which disappears when ${\epsilon_{\rm closed}}$ is analytically continued to $0$, this divergence is a real obstruction and there is nothing we can do about it. \footnote{In a full open-closed SFT, if the closed string is marginally relevant this should trigger a   condensation towards a new closed string background, connected to the initial one by a bulk RG-flow.}
\section{Discussion and outlook}\label{sec:7}
In this paper we have started a systematic study of closed string deformations in Witten-type OSFTs. We have found that the coupling with the Ellwood invariant can  induce a change in the closed string background, which is captured by a vacuum shift solution, an open string state which lives in the original closed string background, with the original boundary conditions. The (perturbative) existence of this vacuum shift solution has been related to the vanishing of S-matrix elements between the deforming closed string and one massless (or physical) open string. These amplitudes represent the tadpole in the effective theory which is obtained by integrating out the open massive (or unphysical) fields. 

Staying in the realm of the bosonic string, there are several directions to explore.
\begin{itemize}
 \item First of all, given the  progress  in the  explicit constructions of analytic solutions  in OSFT,  it is natural  to search for analytic solutions describing the vacuum shift, and in general the shift of any given solution (if this is unobstructed). 

\item The present definition of the Ellwood invariant strictly requires an $h=(0,0)$ primary closed string insertion, otherwise there are midpoint singularities either in Fock space or in star products.\footnote{Divergences in the Fock space can be easily amended by  defining the Ellwood state as $\tilde e=U_1^\dagger V(i,-i)\ket0$. However the star product of this state with generic Fock states is singular at the midpoint (unless $V$ is a $(0,0)$ primary), and this breaks the gauge invariance of the deformation.} Unfortunately this limitation  does not allow to couple the ghost dilaton  $(c\del^2c-\bar c\bar \del^2 \bar c)$ which, although physical, is not primary.  The ghost dilaton is a modulus of the bulk which is expected to control the string coupling constant while not changing the bulk CFT \cite{Bergman:1994qq, Yang:2005ep} and it should couple to open strings. Therefore it is interesting to search for a gauge invariant coupling of the ghost dilaton in Witten OSFT.

\item Despite the great advantage that Witten theory  provides a complete covering of the bosonic moduli space with just open string propagators, the part of the moduli space which is near to closed string degeneration is not naturally regulated.
However, differently from open strings (which are attached to D-branes), closed strings  typically propagate on space-times with non compact directions  and so for generic momenta we can always define the amplitude by analytic continuation, computing it in the convergence region. This is in fact how string theory amplitudes are usually computed. 
Therefore, at generic closed string momenta,  we shall not worry about closed string degeneration. 
However if we are interested in strict zero-momentum closed strings (which is typically the case in background deformations) this method does not appear to work because in general the zero momentum limit of an amplitude depends on how the momenta are sent to zero. What is needed in this case is a universal regularization like \eqref{tame} for the regions $t\sim 1$ of the open string propagators attached to the Ellwood invariants, generalizing our proposal in section \ref{sec:closdeg} for the simplest amplitude $\aver{ e,he}$. It is natural to expect that this regularization (which in general will depend on the given diagram and on the position of the open string propagator inside the diagram) can be obtained by deforming a bit Witten theory to an infinitesimally close-by interpolating open-closed theory \cite{Zwiebach:1992bw}, so that  closed  string products and propagator would be introduced in a gauge invariant way, consistently with a decomposition of moduli space in which closed string degeneration is taken care of by the closed sector.  This is the spirit of the regularization we have used in section \ref{sec:closdeg}.
We hope to come back to this important point soon.
\item What is the cosmological constant \eqref{cosmo} computing? For a perturbative vacuum shift starting linearly in $\mu$ it is formally the resummation of all deforming closed string amplitudes off the disk. It would then be natural to relate this quantity to the shift in the $g$ functions of the D-brane before and after the bulk deformation. However such difference starts at $O(\mu)$ whereas our cosmological constant starts at $O(\mu^2)$. Having a physical quantity to compare it with, would also be helpful for having an independent check  on the regularization we have used in section \ref{sec:closdeg}.
\end{itemize}
We hope to be able to answer the above questions in the near future.

\section*{Acknowledgments}
We thank Ted Erler and Martin Schnabl for discussions and  Ashoke Sen for correspondence and useful comments on a draft of this paper.
We thank the organizers of ``Fundamental Aspects of String Theory'', Sao Paolo 1-12 June 2020,  and in particular Nathan Berkovits for giving us the opportunity to present some of our results prior to publication.
CM thanks CEICO and the Czech Academy of Science for hospitality during part of this work. JV also thanks INFN Turin for their hospitality during the initial stages of this work.  
The work of JV was supported by the Czech Science Foundation - GA\v{C}R under the grant EXPRO 20-25775X.
The work of CM is partially supported by the MIUR PRIN
Contract 2015MP2CX4 ``Non-perturbative Aspects Of Gauge Theories And Strings''.

\appendix
\section{Equivalence of open and closed string exchange}\label{app:2}

In this appendix we show that 
\be
 \langle 0 |\,V_1(i,-i)\frac{b_0}{L_0}V_2(i,-i)\,|0\rangle=-\langle B|\, \frac{b_0^+}{L_0^+} \,V_1(-1,-\bar 1)V_2(1,\bar1)\,|0\rangle,\label{duality}
\ee
by using the conformal map \eqref{map}. We start from the open string side of the above to-be-proven equality
\be
\langle 0| V_1(i,-i)\frac{b_0}{L_0}V_2(i,-i)\,|0\rangle=\int_0^1\frac{dt}{t}\aver{V_1(i,-i)\,b_0\left[V_2(it,-it)\right]}=(*).
\ee
Using $V(z,z^*)=c(z)c(z^*)\VV(z,z^*)$, where $\VV$ is a matter $(1,1)$ primary and 
\be
b_0=\oint\frac{dz}{2\pi i} z b(z),
\ee
we get
\be
(*)&=&i\int_0^1 dt\aver{c(i)c(-i)(c(it)+c(-it))\, \VV_1(i,-i)\VV_2(it,-it)}\\
&=&i\int_0^1 dt\aver{c(i)\bar c(\bar i)(c(it)+\bar c(\overline{it}))\, \VV_1(i,\bar i)\VV_2(it,\overline{it})}_{\rm UHP}=(**),
\ee
where in going from the first to second line we have rewritten the same expression without using the doubling trick. To compare with the boundary state amplitude we map the UHP to the Disk in such a way that the closed strings will be symmetrically inserted around the origin. Up to an unimportant phase this is provided by the $t$-dependent $SL(2,C)$ map
\be
w&=&\frac{\sqrt t+iz}{\sqrt t-iz}\\
\bar w&=&\frac{\sqrt t-i\bar z}{\sqrt t+i\bar z}.
\ee
Performing the elementary holomorphic and anti-holomorphic transformations the correlator becomes
\be
\aver{c(i)\bar c(\bar i)\left(c(it)+\bar c(\overline{it})\right)\, \VV_1(i,\bar i)\VV_2(it,\overline{it})}_{\rm UHP}=\0\\
\frac{(1+x)^3}{2i (1-x)}\aver{c(-x)\bar c(-\bar x)\left(c(x)-\bar c(\bar x)\right)\, \VV_1(-x,-\bar x)\VV_2(x,\bar x)}_{\rm Disk},\label{temp}
\ee
where
\be
x=\frac {1-\sqrt t}{1+\sqrt t}=\bar x.
\ee
Then, changing the integration variable from $t$ to $x$ the Jacobian simplifies with the prefactor in \eqref{temp} and  the full amplitude becomes
\be
(**)&=&2\int_0^1 dx \aver{c(-x)\bar c(-\bar x)\left(c(x)-\bar c(\bar x)\right)\, \VV_1(-x,-\bar x)\VV_2(x,\bar x)}_{\rm Disk}\0\\
&=&2\int_0^1 dx\, \left(-4 x(1-x^4)\right) \aver{ \VV_1(-x,-\bar x)\VV_2(x,\bar x)}_{\rm Disk}\,\label{final-open},
\ee
where we have explicitly evaluated the ghost correlator on the disk.\footnote{This can be mapped to a purely holomorphic correlator on the full complex plane using the gluing condition on the disk $\bar c(\bar z)\to -(z^*)^2 c\left(\frac1{z^*}\right)$.}

We now consider the closed string correlator
\be
&&\langle B|\, \frac{b_0^+}{L_0^+} \,V_1(-1,-\bar1)V_2(1,\bar 1)\,|0\rangle\equiv \aver{\frac{b_0^+}{L_0^+} \,\left(V_1(-1,-\bar1)V_2(1,\bar 1)\right)}_{\rm Disk}\0\\
&=&\int_0^1 \frac{dx}{x}\aver{(b_0+\bar b_0) \,\left(c\bar c\VV_1(-x,-\bar x)c\bar c\VV_2(x,\bar x)\right)}_{\rm Disk}\0\0\\
&=&\int_0^1 \frac{dx}{x}\aver{{\Big (} (x\bar c(\bar x)-x c(x)) c(-x)\bar c(-\bar x)+c(x)\bar c(\bar x) (-x\bar c(-\bar x)+x c(-x)){\Big )}\VV_1(-x,-\bar x)\VV_2(x,\bar x)}\0\\
&=&\int_0^1 dx \left(8 x(1-x^4))\right)\aver{\VV_1(-x,-\bar x)\VV_2(x,\bar x)}_{\rm Disk},
\ee
where in the last line we have computed the 4 ghost correlators (which all give the same contribution). Looking at \eqref{final-open} we see that \eqref{duality} has been proven.

\endgroup
\end{document}